\documentclass[aip,amsmath,amssymb,reprint]{revtex4-1}

\usepackage{graphicx}
\usepackage{xcolor}
\usepackage{dcolumn}
\usepackage{bm}
\usepackage{bbm}
\usepackage[normalem]{ulem}

\usepackage[utf8]{inputenc}
\usepackage[T1]{fontenc}
\usepackage{mathptmx}

\usepackage[colorinlistoftodos,prependcaption,textsize=tiny]{todonotes}
\usepackage{multirow}

\usepackage{pifont}
\newcommand{\cmark}{\ding{51}}%
\newcommand{\xmark}{\ding{55}}%

\usepackage{bbold}
\usepackage{float}

\usepackage{natbib}

\def \kram {{\mathcal{K}}}
\def \kramhat {{\hat{\kram}}}
\def \imag {{\text{i}}}
\def \bfj {{\bf j}}
\def \bfi {{\bf i}}

\def \Di {{D_{\bfi}}}
\def \Ci {{C_{\bfi}}}
\def \Cibar {{C_{\bar{\bfi}}}}
\def \Dj {{D_{\bfj}}}
\def \Cj {{C_{\bfj}}}

\def \braDi {{\langle \Di | }}

\def \ketDi {{| \Di \rangle }}
\def \braPsi {{\langle \Psi |}}
\def \ketPsi {{| \Psi \rangle }}

\def \ketPsihf {{| \Psi_\text{HF} \rangle }}

\def \half {{\frac{1}{2}}}
\def \Eh {{\text{E}_h}}

\newcommand{\cre}[1]{ \hat{a}_{#1}^{\dagger} }

\newcommand{\des}[1]{ \hat{a}_{#1} }

\newcommand{\rndbrk}[1]{ \left({#1}\right)}

\begin{document}
\author{Robert J. Anderson}
\affiliation{Department of Physics, King's College London, Strand, London, WC2R 2LS, U.K.}
\author{George H. Booth}
\email{george.booth@kcl.ac.uk}
\affiliation{Department of Physics, King's College London, Strand, London, WC2R 2LS, U.K.}

\title{Four-component Full Configuration Interaction Quantum Monte Carlo for Relativistic Correlated Electron Problems}

\begin{abstract}
An adaptation of the full configuration interaction quantum Monte Carlo (FCIQMC) method is presented, for correlated electron problems containing heavy elements and the presence of significant relativistic effects. The modified algorithm allows for the sampling of the four-component spinors of the Dirac--Coulomb(--Breit) Hamiltonian within the relativistic no-pair approximation. 
The loss of spin symmetry and the general requirement for complex-valued Hamiltonian matrix elements are the most immediate considerations in expanding the scope of FCIQMC into the relativistic domain, and the alternatives for their efficient implementation are motivated and demonstrated.
For the canonical correlated four-component chemical benchmark application of Thallium Hydride, we show that the necessary modifications do not particularly adversely affect the convergence of the systematic (initiator) error to the exact correlation energy for FCIQMC calculations, which is primarily dictated by the sparsity of the wave function, allowing the computational effort to somewhat bypass the formal increases in Hilbert space dimension for these problems. We apply the method to the larger problem of the spectroscopic constants of Tin Oxide, correlating 28 electrons in 122 Kramers-paired spinors, finding good agreement with experimental and prior theoretical relativistic studies.
\end{abstract}
\maketitle

\section{Introduction}
Relativistic effects play an important role in chemistry and physics where heavier elements are present, including fundamental changes in both structural and spectroscopic properties \cite{doi:10.1063/1.3702628, doi:10.1146/annurev-physchem-032511-143755,Knecht2019}. Classic examples such as the colour of gold have been found to be more widely representative of qualitative changes due to effects originating from a relativistic treatment. These effects manifest across a wide domain of applications, such as the contraction of core orbitals in shifting the NMR constants of heavy elements\cite{doi:10.1021/ct400726y}, or the sensitivity of dynamics and spectroscopy to the treatment of spin-orbit coupling phenomena\cite{ISMAIL1999296}. Any computational electronic structure method must therefore be able to consider these effects for it to claim a broad scope of applicability in treating these chemically important systems. 

Furthermore, the importance of an accurate treatment of correlation amongst many valence electrons is well-understood for obtaining qualitatively correct chemical insights. 
However, the combination of strong correlation effects with large relativistic contributions remains a significant challenge in electronic structure, which occur in many physical systems \cite{doi:10.1063/1.471544}. For instance, numerical treatment of low-spin or charge transfer excited states even in light-element molecules, is another case in which correlated wave function expansions are required for an adequate characterisation, and one in which effective magnetic interactions from a relativistic picture can be a vital inclusion, especially in the study of spin-forbidden transitions\cite{doi:10.1021/acs.chemrev.7b00060} and conical intersections \cite{doi:10.1063/1.1378324}. Therefore, the development of methodology and software capable of solving the relativistic many-body problem for large systems is clearly a necessary endeavour.

Strong (or `static') correlation effects denote a significant entanglement of many electrons beyond that captured in a single-particle description, and is often expensive to tackle computationally given the need for large expansions of the many-body wave function. In the absence of relativistic effects, a number of methods have come to the fore in recent years which can in principle tackle strong molecular correlation in a systematically improvable fashion. These include the density matrix renormalization group \cite{white1992} (DMRG), a resurgence in selected configuration interaction methods\cite{sharma2017} (SCI), and Fock space stochastic methods such as full configuration interaction quantum Monte Carlo (FCIQMC). Recently, DMRG\cite{doi:10.1021/acs.jctc.7b01065} and SCI\cite{sharma2018} have been combined with a relativistic treatment to different levels, as well as rigorous four-component approaches also devised for more traditional single- and multi-reference quantum chemical methods for strong correlation, such as coupled-cluster, complete active space self-consistent field and multireference perturbation theories\cite{shiozaki2015,doi:10.1021/acs.jctc.8b00910,doi:10.1021/acs.jctc.9b01290,doi:10.1063/1.4906344}. It is a systematically improvable treatment of strongly correlated physics within a rigorous four-component formalism which we aim to address in this work, by adapting the successful FCIQMC framework for correlated systems to be amenable to sampling in this space of spinors of the relativistic Dirac equation.


A number of different approaches have been developed to make the Dirac theory of the relativistic electron generally compatible with many-body quantum chemistry\cite{doi:10.1063/1.473860, doi:10.1063/1.472460,doi:10.1063/1.3159445}.
Most difficulties with this undertaking stem from the property that the Hamiltonian is not bounded from below, formally invalidating standard techniques that rely on the existence of a variational principle.
The construction of a small component basis in ``kinetic balance''\cite{doi:10.1002/qua.560250112} with that of the large component proves to be sufficient in maintaining the correct non-relativistic limit relationship between the two bispinors, making it possible to distinguish between the positive energy solutions and undesired negative energy solutions in a mean field calculation, denoted {\em Dirac}--Hartree--Fock (DHF) \cite{shiozaki2013}.

Moving to the interacting many-body problem, the continuum dissolution pathology of the four-component Dirac Hamiltonian manifests in another formal problem which would render a system of two interacting electrons unstable, since transitions to arbitrarily excited positive and negative energy continuum states are allowed.
This is commonly cured by embedding the resulting Hamiltonian within projection operators which cause matrix elements involving integrals over positronic single-particle functions to vanish. This four-component \emph{no virtual pair approximation} (4c-NVPA) \cite{doi:10.1063/1.4959452} is built upon a second-quantised formulation of the many-body problem, and so Fock-space based methods become not merely convenient, but mandatory in this context.
The computation of matrix elements becomes more involved in the move from non-relativistic Schr{\" o}dinger theory to fully relativistic theory, as the integrals over products of scalar functions are replaced by inner products of four-spinors. Ultimately however, the general second-quantised 4c-NVPA Hamiltonian is parametrisable by arrays of complex-valued one- and two-electron matrix elements in a similar fashion to the non-relativistic theory, as

\begin{equation}
\label{eq:ham}
    \hat{H}_\text{NVPA} \equiv 
    \sum_{ij} h_{ij}
    \cre{i}\des{j} 
    +
\half{}
    \sum_{ijkl} (g_{ijkl}-g_{ijlk})
    \cre{i}\cre{j}\des{l}\des{k}.
\end{equation}

The $h$ and $g$ arrays are the one- and two-body integrals:
\begin{equation}
    h_{ij} \equiv \int \mathrm{d}^3 (1) 
    \phi_i(1)^\dagger
    \hat{P}^{+}_{1}
    \hat{h}(1)
    \hat{P}^{+}_{1}
    \phi_j(1)
\end{equation}
\begin{equation}
    g_{ijkl} \equiv \int \mathrm{d}^3(1)\,\mathrm{d}^3(2)\, 
    \phi_i(1)^\dagger
    \phi_j(2)^\dagger
    \hat{P}^{+}_{1}
    \hat{P}^{+}_{2}
    \hat{g}(1, 2)
    \hat{P}^{+}_{1}
    \hat{P}^{+}_{2}
    \phi_k(1)
    \phi_l(2)
\end{equation}
where $\phi_i$ are four-component spinors, and $\hat{P}^{+}_{\vec{x}}$ are the projectors into the space of electronic single particle wavefunctions. In this work, $\hat{h}$ stands for the full Dirac one-body Hamiltonian, fully including spin-orbit, contact and scalar relativistic contributions to all orders, and $\hat{g}$ is the Coulomb--Breit operator, including relativistic magnetic and retardation effects between the electrons, correct up to $\mathcal{O}[1/c^2]$ \cite{doi:10.1063/1.4795430}. This Hamiltonian is often called the Dirac-Coulomb-Breit (DCB) Hamiltonian.
Fock-space formulated correlation methods therefore can be made compatible with two- or four-component relativistic Hamiltonians with generally only minor algorithmic modifications \cite{C6CP07588F,doi:10.1021/jp904914m,doi:10.1063/1.4862495}. The aim of the present work is to investigate whether the required modifications to the initiator FCIQMC algorithm result in an efficient stochastic methodology for systems in which kinematic and magnetic relativistic effects have an appreciable effect on electron correlation. 

\section{Brief Recap of FCIQMC} 
The ground state solution of the imaginary time Schr{\" o}dinger equation can be expressed as a product of discrete linear propagation steps with a small timestep $\tau$, as
\begin{equation}
    \ketPsi = \lim_{N\rightarrow\infty}(1-\hat{H}\tau)^N\ketPsihf,
    \label{eq:determ_master}
\end{equation}
where the CI vector is $\ketPsi \equiv \sum_\bfi \Ci \ketDi$, with $\bfi$ encompassing the set of all possible combinations of occupied single particle states within the full Hilbert space of electronic configurations.
This configurational (FCI) space grows exponentially with system size, and so efforts to deterministically iterate Eq. \ref{eq:determ_master} meet with computational impracticality for even relatively small numbers of electrons and correlated degrees of freedom.
The FCIQMC method \cite{doi:10.1063/1.3193710} delays the onset of this exponential wall to access larger system sizes with high accuracy, by representing each CI coefficient as a population of discrete ``walkers'', and recasting Eq.~\ref{eq:determ_master} as a stochastic master equation governing their dynamics. The observation that the CI vector is often sparse given a suitably-chosen single particle basis is central to the method's efficacy. The algorithm effectively compresses the CI vector following each iteration of Eq.~\ref{eq:determ_master} via an unbiased stochastic rounding of each non-zero amplitude, such that a large number of configurations can be removed from consideration in the subsequent step.

A single application of the propagator $(1-\hat{H}\tau)$ results in the following update to each coefficient:
\begin{equation}
    \Delta \Ci = \overbrace{-\tau\sum_{\bfj \neq \bfi}  H_{\bfi\bfj}\Cj}^\text{off-diagonal (spawning)}
    \overbrace{-\tau(H_{\bfi\bfi}-S_0)\Ci}^\text{diagonal (death)}.
    \label{eq:fciqmc_master}
\end{equation}
In the semi-stochastic FCIQMC algorithm\cite{PhysRevLett.109.230201}, the first summation (spawning) is split into a subspace, such that all ${\bf i, j}$ pairs within the subspace are propagated exactly and deterministically, while all other pairs have their off-diagonal contributions from $\bfi \leftarrow \bfj$ made stochastically via a spawning step, ensuring that these contributions from the Hamiltonian are made sparsely.
The number of attempts to spawn walkers from $\Dj$ is determined by a stochastic round of the instantaneous $\Cj$, so that in the original algorithm, the $\Delta \Ci$ due from a successful excitation generation is
\begin{equation}
    -\frac{\tau H_{\bfi\bfj}}{p_\text{gen}(\bfj\leftarrow\bfi)}\frac{\Ci}{n_\text{attempt}(\Ci)}.
    \label{eq:spawning_update}
\end{equation}

In the initiator scheme---where an initiator determinant is defined as one occupied by more than $n_\text{add}$ walkers---this contribution is only made if $\Di$ is found to be already occupied by a walker, or $\Dj$ is an initiator, or a simultaneous contribution is made in a single iteration from another non-initiator determinant. This adaptation ($i$-FCIQMC) eases the propagation of sign-incoherence noise due to small occupations, but introduces a systematic error which is only eliminated in the limit of a large number of walkers $N_W$. This adaptation is made throughout this work, with $n_\text{add}=3.0$.
The diagonal term in Eq. \ref{eq:fciqmc_master} simply scales $\Ci$ if $\Di$ is in the deterministic subspace, otherwise the $\tau(H_{\bfi\bfi}-S_0)$ factor is interpreted as a death rate at which walkers are to be stochastically removed, where $S_0$ is a diagonal shift which is modulated to maintain the walker population at the desired level.

\section{Relativistic FCIQMC}
Due to the coupling of spin and spatial parts of the Dirac Hamiltonian, neither the $\hat{S}_z$ nor $\hat{S}^2$ operators commute with the Dirac Hamiltonian, annulling their use as good quantum numbers for restricting the FCIQMC dynamics. However, time reversal remains a symmetry of the system, and can in some sense take over the role of a discrete symmetry analogously to $\hat{S}_z$ with which the construction of the single particle basis can be restricted. Applying this restriction at the mean field level results in a Kramers-paired basis\cite{doi:10.1063/5.0015279}, in which a spinor $\phi_i$ is related to its Kramers partner $\bar{\phi}_i$ by the application of the antilinear time reversal operator
\begin{equation}
    \kramhat = -\imag 
    \begin{pmatrix}
\sigma_y & 0_2 \\
0_2 & \sigma_y 
\end{pmatrix}
    \kramhat_0
\end{equation}
where the perpendicular Pauli matrices have the effect of flipping the spins (and changing sign when acting on the barred spinor) of the large and small bispinors, and the complex conjugation operator $\kramhat_0$ implies time reversal since the time variable appears multiplied by the imaginary unit in the Dirac equation. This operator relates Kramers partners $\phi_i$ and $\bar{\phi}_i$ by:
\begin{equation}
    \kramhat\phi_i \equiv \bar{\phi}_i ;\quad \kramhat\bar{\phi}_i \equiv -\phi_i
\end{equation}
The relativistic analogue of the many-body $z$-projected spin $M_S$ is therefore defined as
\begin{equation}
M_K\equiv \frac{1}{2}\sum_p \braPsi \cre{p}\des{p} - \cre{\bar{p}}\des{\bar{p}} \ketPsi.
\end{equation}
Spatial point-group symmetries cannot be directly applied to molecular systems in a four-component formalism, since a rotation of $2\pi$ results in a change of sign for fermion functions. Instead they must be augmented by the inclusion of a new element $\bar{E}$, representing a complete rotation about an arbitrary axis. These are known as \emph{double groups} owing to a doubling in the number of symmetry operations to include the effect of time-reversal. The additional irreducible representations are dubbed \emph{fermion irreps}, while those spanning the original point group are called \emph{boson irreps}. Typographically, double groups are distinguished from their bosonic counterparts with the addition of a superscript $*$.

The implementation of quaternion algebra is advantageous in high symmetry situations to fulfil these double-group symmetries \cite{doi:10.1063/1.479958}. As shown in the first two rows of Table \ref{tab:double_groups}, the combinations of Kramers barred and unbarred spinors which may appear in symmetry-allowed non-zero integrals can be restricted by the imposition of applicable double group symmetries, increasing the sparsity of the Hamiltonian matrix and restricting the size of the accessible CI space. In this paper, we adapt the FCIQMC algorithm for sampling configurations constructed from four-component spinors, and investigate the effect that symmetries in this formulation can have on FCIQMC efficiency, including reduction to real-valued matrix elements and reduction in Hilbert space dimension within a four-component relativistic formulation.
\def\arraystretch{1.5}
\begin{table}[]
\begin{tabular}{c|c|c|c|c}
\hline
symmetry                          & $h_{\bar{p}q}{=}0$ & $g_{\bar{p}rqs}{=}0$ & $g_{\bar{p}\bar{r}qs}{=}0$ & real-valued \\
\hline
\multicolumn{5}{c}{Kramers-restricted molecular spinors} \\
$D_{2h}^*$, $D_{2}^*$, $C_{2v}^*$ & \cmark                                        & \cmark                           & \xmark                                 & \cmark      \\
$C_{2h}^*$, $C_{2}^*$, $C_{s}^*$  & \cmark                                        & \cmark                           & \xmark                                 & \xmark      \\
$C_{i}^*$, $C_{1}^*$              & \cmark                                        & \xmark                           & \xmark                                 & \xmark      \\
\hline
\multicolumn{5}{c}{Kramers-unrestricted molecular spinors} \\
None                              & \xmark                                        & \xmark                           & \xmark                                 & \xmark     \\
\hline
\end{tabular}
\caption{
\label{tab:double_groups}
Constraints on the value and type of matrix elements arising from a consideration of time-reversal and point group symmetry. Barred spinor indices indicate the time-reversed spinor, showing that the two-body interaction can indeed break $M_K$. Also indicated is whether non-zero integrals as defined in Eq. \ref{eq:ham} can be expressed without an imaginary part, determined by the double group to which the which the system in question belongs.
}
\end{table}

The level of code modification required beyond standard non-relativistic (or scalar relativistic) FCIQMC in order to apply the method to four-component relativistic correlation problems depends upon the symmetries respected in the preparation of the molecular spinor basis. We have developed the most general implementation, by generalising the precomputed heat bath excitation generator \cite{doi:10.1021/acs.jctc.8b00844,
doi:10.1021/acs.jctc.5b01170, doi:10.1063/5.0005754} to support all orders of $M_K$ non-conservation, and adapting the entire algorithm for complex-valued walkers.
This latter consideration has been partially dealt with in Ref. \onlinecite{nature_booth} in the consideration of $k$-point symmetry for translationally symmetric systems. In this, the master equation was broken down into its real and imaginary components, and real and imaginary walkers propagated as different `types'. However, the concept of real and imaginary components of the stochastic CI coefficients corresponding to two separate but inter-spawning walker populations is an unnecessary artefact. As such, quantities such as the number of spawning attempts, total walker number and the initiator criteria were evaluated based separately on the magnitude of the real and imaginary component of a determinant occupation, rather than the magnitude of the complex occupation.

In this work, we have developed support for manifestly complex walker populations, and the elimination of the need to consider cross-spawning between real and imaginary populations, thereby reducing communication overhead for complex-valued stochastic wavefunctions. The number of complex walkers $N_W$ on a determinant $\Di$ is now taken to be $\sqrt{\mathfrak{R}\Ci^2+\mathfrak{I}\Ci^2}$ in contrast to the previously used $|\mathfrak{R}\Ci|+|\mathfrak{I}\Ci|$. Thus, the newly spawned walker magnitudes defined in Eq. \ref{eq:spawning_update} are given a complex phase by the parent walker, rather than a factor of a signed real or imaginary unit. This simple change to the complex walker algorithm also means that initiator status of an occupied determinant is no longer determined for the real and imaginary components separately. Instead it is decided for the determinant as a whole in a manner much more closely analogous to the real walker $i$-FCIQMC algorithm.

The issue of $M_K$ non-conservation of the walker dynamic as detailed in Tab.~\ref{tab:double_groups} is effectively addressed with the emergence of the memory-bound but highly computationally efficient heatbath random excitation generators for the spawning step\cite{doi:10.1021/acs.jctc.5b01170,doi:10.1063/5.0005754}. This algorithm results in a very strongly-peaked $H_{\bfi\bfj}/p_\mathrm{gen}(\bfj \leftarrow \bfi)$ distribution, and crucially does not require special consideration of symmetries.
The complex, $M_K$ non-conserving heatbath excitation generators developed for this investigation precompute a $N_\mathrm{orb}{}^4$-scaling array of probabilities which can be probed in constant time via the aliasing method.
These excitation sampling algorithms are set up for the double excitations alone, whose magnitudes depend only on the orbitals which differ in the bra and ket determinants they connect; singles on the other hand are rare compared to doubles, and since their magnitudes depend on the orbitals occupied in common between the bra and ket, they are generated uniformly.
Aside from greatly increasing the maximum stable value for the timestep, these excitation generators innately conserve the symmetries of the underlying sampled Hamiltonian.
This fortunate property of the excitation generator allows us to disregard the symmetry labels, since excitations forbidden by a symmetry of the Hamiltonian are never drawn.

Symmetry considerations also do not play a part in the data layout, since in an efficient FCIQMC implementation there is no fixed layout in memory for the determinant space, which is instead built-up in an associative array through symmetrically-conservative spawning.
Determinant-wise time-reversal symmetry is neglected at the present time, but it will be a necessary future addition to the algorithm in order to target a $\kramhat$-antisymmetric eigenfunction by evolution to the ground-state, in addition to conferring an approximate halving of the $N_W$ required to converge energy estimators and other expectation values. Since no relationship between $\Ci\equiv\braDi\Psi\rangle$ and $\Cibar\equiv\braDi\kramhat_N|\Psi\rangle$ is enforced, where $\kramhat_N=\otimes_i \kramhat_i$ is the many-particle time-reversal operator, the present four-component FCIQMC algorithm is said to be \emph{Kramers-unrestricted}.

\section{Results}
\subsection{The additional cost of 4c-FCIQMC}
Due to the use of a more complicated algebra, coupled with the loss of spin symmetry, four-component CI problems are generally accompanied by a substantial growth in the Hilbert space dimension and computational cost \cite{molfdir, luciarel}. This is problematic for exact, string-based CI solvers, however the present investigation aims to determine whether the inherent sparsity of the stochastic wavefunction representation in FCIQMC is able to offset this additional cost.
We consider a Thallium Hydride (TlH) system at a bondlength of 1.872\AA, a standard benchmark molecule in the relativistic quantum chemistry literature \cite{3591c35695224fa4a56e6cae18a25490, doi:10.1063/1.1389289}. A 14 electron, 47 Kramers pair (94 spinor) active space was selected from DHF orbitals in a Dyall cv3z basis set, into which the Dirac--Coulomb--Breit Hamiltonian was projected. This utilized the DMRG interface of the DIRAC code\cite{DIRAC13,doi:10.1063/5.0004844}.

\begin{figure}[t]
\centering
\includegraphics[width=0.45\textwidth]{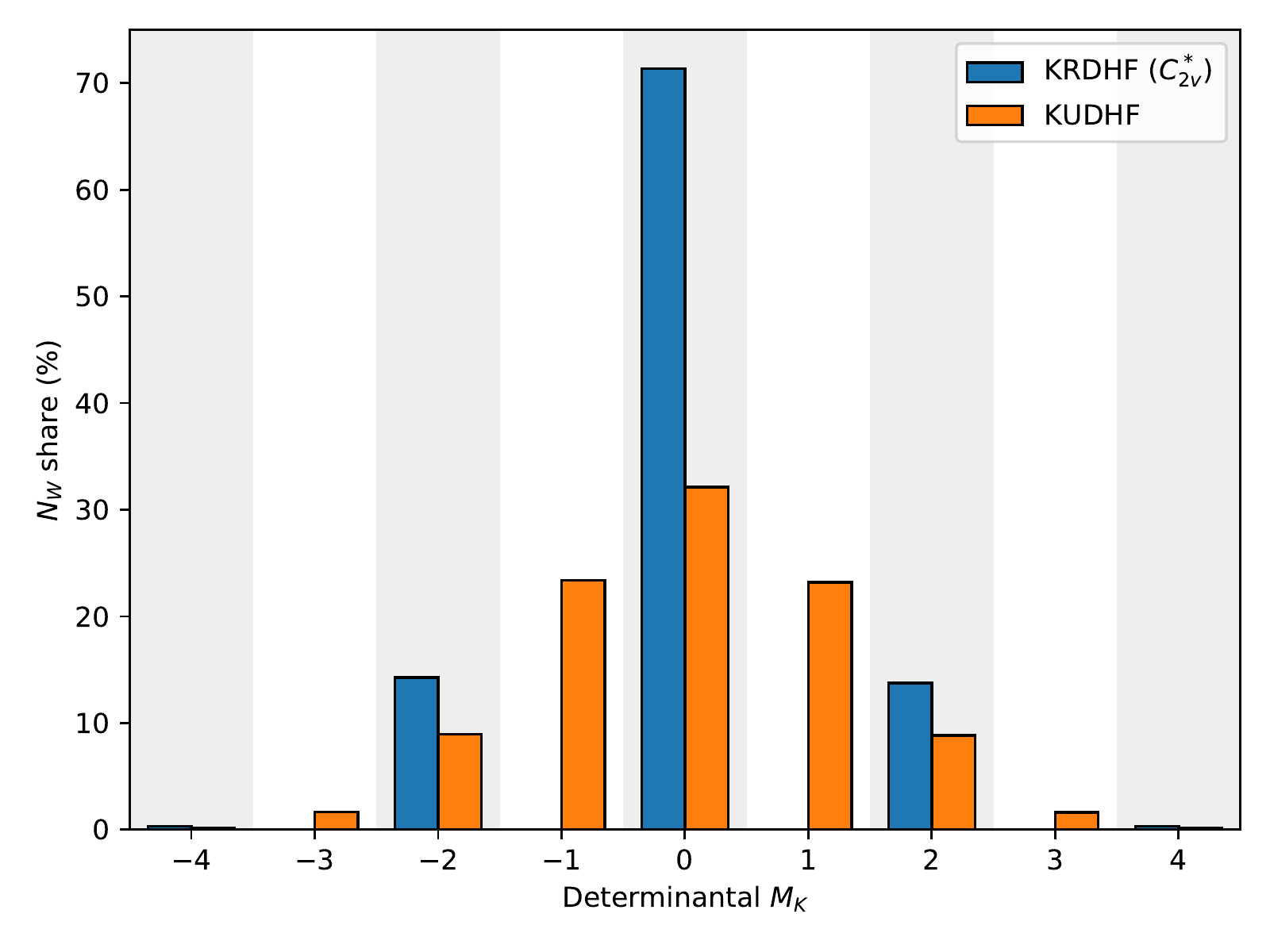}
	\caption{
\label{fig:mk_nw}
	Fraction of walkers residing in each $M_K$ sector of the Hilbert space for an equilibrated FCIQMC calculation with $50 \times 10^6$ walkers for TlH in a 14 electron, 94 spinor active space. KRDHF indicates that a Kramers-restricted single-particle DHF basis is constructed, while KUDHF indicates a Kramers-unrestricted basis with complex walker populations.
	}
\end{figure}

In particular, for TlH a $C_{2v}^*$ double point group is employed. In this scheme, two-body integrals are real-valued, and only those of the $g_{pqrs}$, $g_{\bar{p}\bar{q}rs}$, and $g_{pq\bar{r}\bar{s}}$ types can be non-zero.
This implies that $M_K$ non-conservation only appears as simultaneous flipping of two unbarred occupied spinors to two barred spinors, or vice versa, in the FCIQMC walker dynamic, changing $M_K$ by two. Consequently, the CI coefficients are identically zero in odd $M_K$ sectors - resulting in a more compact walker population as shown in Figure \ref{fig:mk_nw}.
 
The figure also shows the (complex) walker distribution when no symmetries are exploited in the preparation of a Kramers-unrestricted Dirac--Hartree--Fock (KUDHF) basis, leading to a full combination of all possible $M_K$ breaking in the Hamiltonian. Apart from the added expense of complex arithmetic, it is clear that the Kramers-unrestricted basis leads to a walkers population which has significant fractions in other $M_K$ sectors of the Hilbert space.
The approximate many body Hilbert space dimension in this case of a KUDHF basis is $1.7\times 10^{16}$ configurations to sample, as opposed to $5.7\times 10^{15}$ for the double group-adapted Kramers-restricted (KRDHF) spinors.
However, this doesn't necessarily directly translate into a corresponding loss of sparsity in the state, for which we analyse the more important metric of convergence of the energy of the state with walker number, which governs the computational cost of an FCIQMC calculation.

As is standard in FCIQMC calculations, the walker population is grown from a small initial occupation on the reference DHF determinant until it reaches a predetermined $N_W$. At this point, the diagonal shift begins to vary so as to maintain the walker population around that level. As the walker distribution begins to equilibrate about the ground state wavefunction, the semi-stochastic adaptation is activated. In these calculations, the set of all occupied single and double excitations of the reference determinant is selected to comprise the deterministic subspace. Semi-stochastic evolution then continues until the trial wavefunction projected energy estimator \cite{doi:10.1063/1.5037923} (in this case, the trial wavefunction is just the reference determinant) is stabilised about a mean value for a sufficient number of iterations to allow a reliable estimation of the standard error from a blocking analysis\cite{doi:10.1063/1.457480}.

\begin{figure}[t]
\centering
\includegraphics[width=0.5\textwidth]{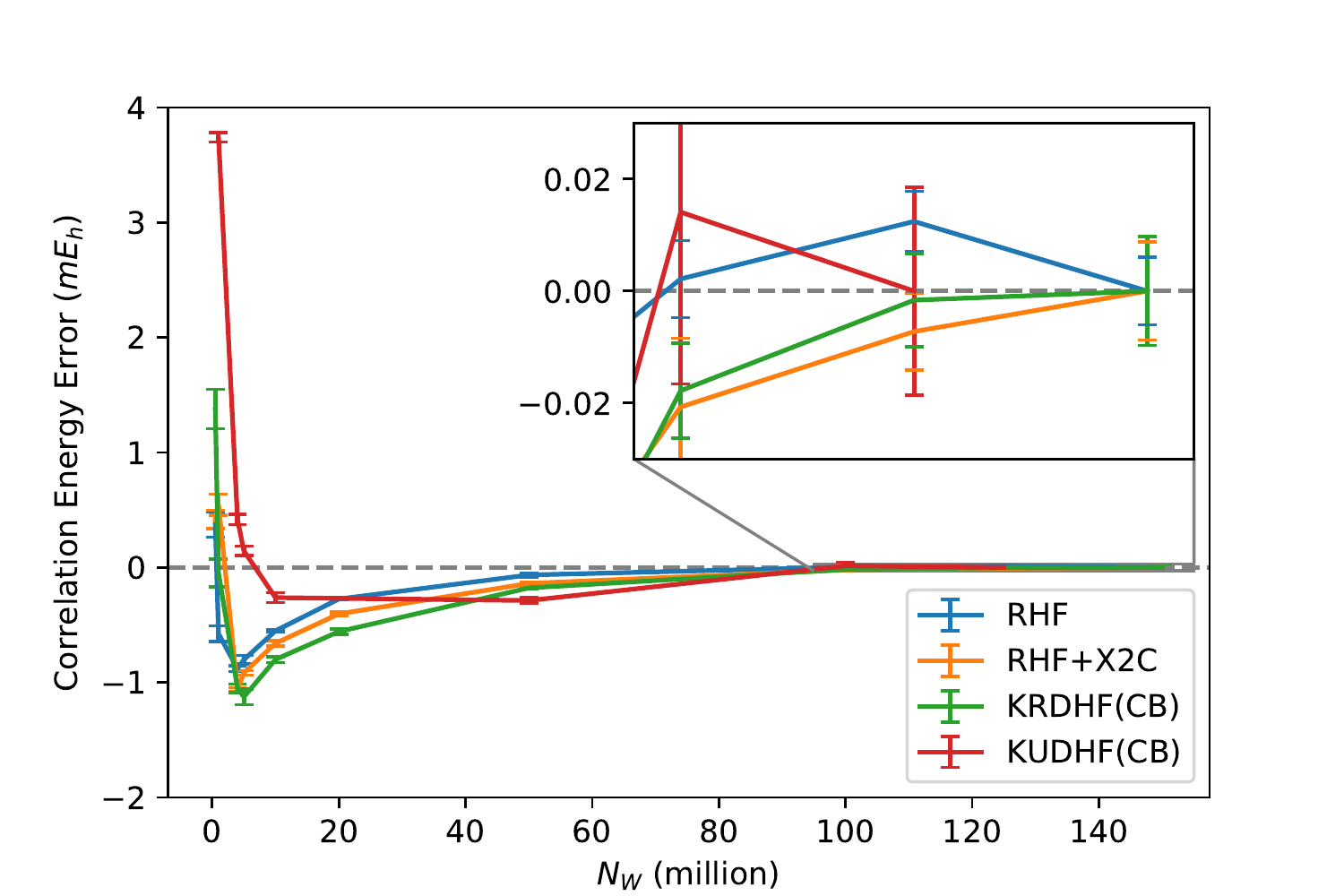}
	\caption{
		Convergence of initiator error in the FCIQMC correlation energy (in $mE_{\mathrm h}$) against total walker number for TlH in a 14 electron, 47 Kramers pair (94 spinor) active space, selected from canonical HF orbitals in the Dyall cv3z basis. The non-relativistic (Schr{\" o}dinger), and spinfree eXact 2-Component hamiltonians are treated with the standard FCIQMC methods, while in the case of the 4-component Dirac--Coulomb--Breit (4c-DCB) Hamiltonian, molecular spinor bases were prepared with Dirac--Hartree--Fock in Kramers-restricted (KRDHF) and -unrestricted (KUDHF) basis representations. The Kramers-restricted basis is constrained to transform in a $C_{2v}{}^*$ representation from the DIRAC package\cite{DIRAC13}, while the Kramers-unrestricted basis is non-symmetry adapted from the BAGEL package \cite{BAGEL}.
		The spinfree X2C Hamiltonian was prepared in the PySCF package\cite{PYSCF,doi:10.1063/5.0006074}.
		The correlation energy is defined as the difference between the HF energy and the mean projected energy estimator at $150 \times 10^6$ walkers, at which the estimator is converged within random error bars with respect to walkers for all Hamiltonians as illustrated in the inset.
	}
\label{fig:TlH_initiator}
\end{figure}

The systematic saturation of initiator error in \mbox{$i$-FCIQMC} for non-relativistic, spinfree X2C\cite{doi:10.1063/1.3159445}, and DCB Hamiltonians with respect to $N_W$ is shown in Fig.~\ref{fig:TlH_initiator}. This convergence profile is encouraging evidence that despite the formal increase in the size of the accessible Hilbert space relative to the non-relativistic and spinfree relativity Hamiltonians (as shown in Figure \ref{fig:mk_nw}), in practice there is no commensurate increase in the number of walkers---and hence, the computational expenditure---required to converge the projected energy estimator within tight random errors of $10^{-5}\Eh$ for the four component Hamiltonians.
Furthermore, the saturation of initiator error with respect to $N_W$ is not substantially slower for the non-symmetry adapted KUDHF orbitals than it is for the $C_{2v}{}^*$-adapted KRDHF orbitals, although the profile of the convergence is slightly modified.
We observe that due to efficient exploitation of wavefunction sparsity by the FCIQMC algorithm, there is relatively small penalty associated with treating these systems without mean-field level symmetry adaptation. Rather than a substantial increase in the walker number, the small additional algorithmic costs in the use of a KUDHF basis primarily stem from the use of complex arithmetic and the timestep alteration due to the additional $M_K$-breaking Hamiltonian terms which must be sampled in the dynamic.

As a means of validating the accuracy and systematic improvability of the 4c-FCIQMC method, calculations were converged with respect to walker number using an identical Hamiltonian to that of TlH at the experimental bond length of $1.872\AA$ from a previous benchmark study\cite{doi:10.1063/1.4862495}, and compared to other high-level quantum chemical four-component implementations.
As opposed to the results of Fig.~\ref{fig:TlH_initiator}, the orbital space is constructed in the basis of 4c-MP2 natural orbitals, which are likely to give a more compact and realistic active space for the calculations compared to a Hartree--Fock energy window active space construction in the previous example. All spinors with an MP2 natural occupation number of between 0.001 and 1.98 are included, giving an active space of 14 electrons in 94 spinors. The integrals defining the arrays $h$ and $g$ in Eq. \ref{eq:ham} are identical to those prepared by the authors of the referenced paper, in which the Kramers restricted molecular spinors are constructed in $C_{2v}{}^*$ symmetry from Dyall cv3z (Tl) and cc-pVTZ (H) atomic basis sets.

\begin{figure}[t]
\centering
\includegraphics[width=0.5\textwidth]{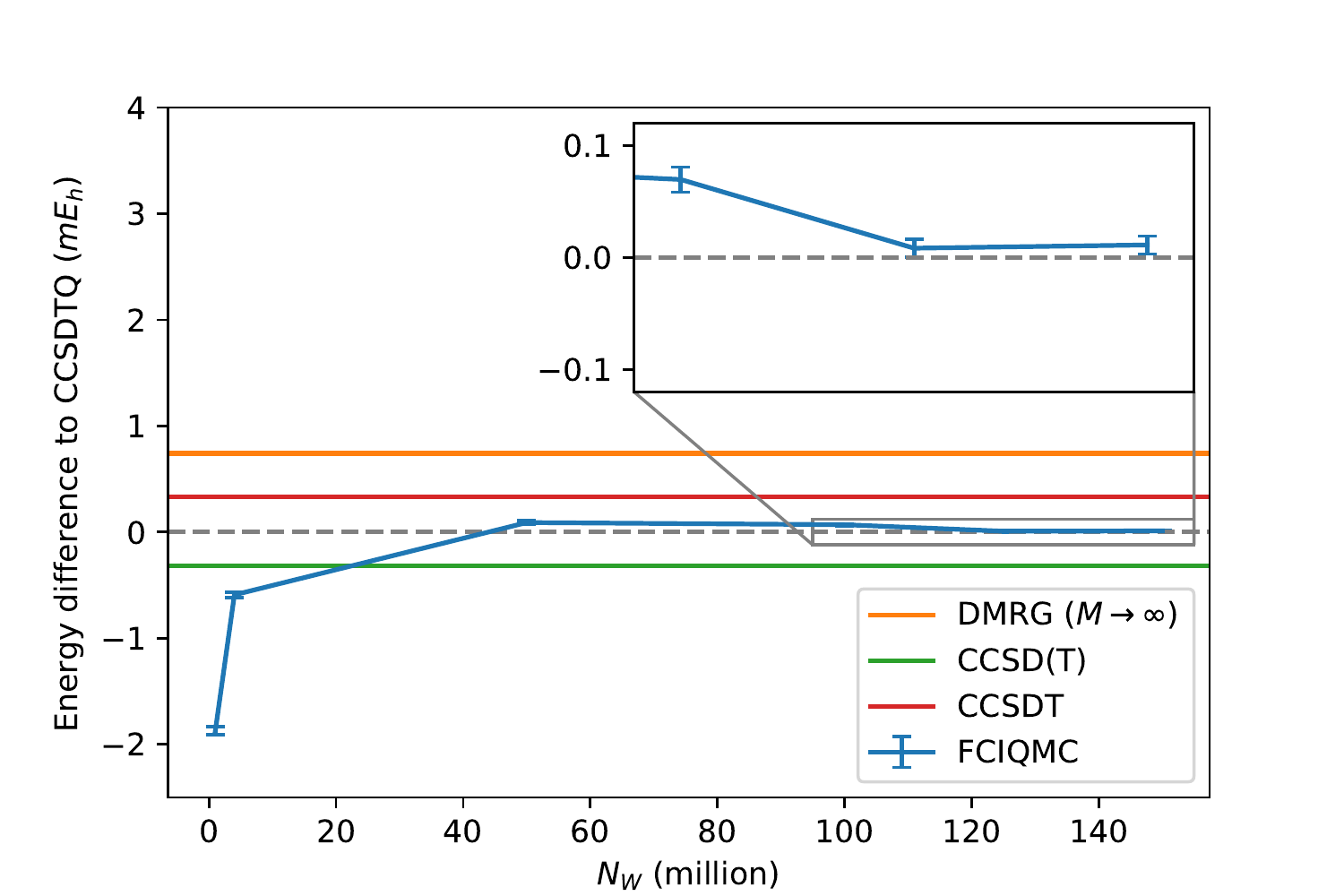}
	\caption{
		Convergence of 4c-FCIQMC energy (in $mE_{\mathrm h}$) relative to 4c-CCSDTQ for the TlH system reported in Ref. \onlinecite{doi:10.1063/1.4862495}, correlating 14 electrons in 94 spinors. Also included for comparison are 4c-DMRG energies (extrapolated from a maximum bond dimension of 4500), and two more truncated coupled cluster results taken from Table 1 of the same publication.
	}
\label{fig:TlH_initiator_knecht}
\end{figure}

Figure \ref{fig:TlH_initiator_knecht} shows the convergence of the 4c-FCIQMC projected energy estimator with respect to total walker population. The result is a mean energy estimate which agrees with a 4c-CCSDTQ calculation to within $\sim2$cm${}^{-1}$. Statistically significant deviation from these results is found for 4c-CCSDT, 4c-CCSD(T) and 4c-DMRG results extrapolated from a maximum bond dimension of 4500, although all these results can be considered to be within chemical accuracy of the benchmark 4c-FCIQMC results.

The electronic structure of TlH is known to be dominated by dynamic correlation, and as such the high-rank coupled-cluster methods are expected to perform very well. However, this does mean that methods which excel in their application to statically correlation, such as DMRG and FCIQMC, are only likely to achieve good comparison to experiment upon expansion to a large basis set, which is unnecessarily costly for these methods. Recent developments have coupled active space 4c-DMRG calculations with large basis corrections, via multi-reference perturbation theory\cite{shiozaki2015, doi:10.1021/acs.jctc.8b00910}, range-separated DFT\cite{doi:10.1063/1.4922295} and tailored coupled cluster approaches\cite{doi:10.1063/1.5144974,doi:10.1063/1.2000251, doi:10.1063/1.2180775}. While this has been performed for FCIQMC in a non-relativistic framework previously, this is a clear research direction to pursue\cite{doi:10.1063/1.5140086,doi:10.1080/00268976.2020.1802072}.
However in the present investigation, four-component FCIQMC energies within active spaces of 94 spinors are found to have converged to within chemical accuracy by only a few million walkers, which constitutes an entirely routine calculation. This corroboration of 4c-FCIQMC via comparison to other highly accurate methods confirms the ability of FCIQMC to produce highly-accurate energy estimates even in a relativistic context.

\subsection{Application to SnO dissociation}
Tin monoxide is an example of a molecular system in which both electronic correlation and relativity play a role in obtaining quantitatively accurate spectroscopic values, and so it is a suitable target application for the relativistic FCIQMC method \cite{doi:10.1063/1.1690757}.
From a Kramers-restricted Dirac Hartree--Fock calculation using the Dyall v4z basis set, occupied orbitals are frozen below -6$\Eh$, while virtual orbitals are truncated above 2$\Eh$. A significant gap in the DHF spectrum is maintained across small geometric distortions about equilibrium for this energy window, and large internuclear distances were not attempted. The resulting correlation problem consists of 28 correlated electrons in 122 Kramers-paired molecular spinors, well beyond the capabilities of a traditional diagonalization approach. The walker populations were grown to $2\times 10^8$, before being stabilised with the onset of variable shift. The semi-stochastic space is instigated after the population has begun to equilibrate, by selecting all determinants with an instantaneous weight larger than $8n_\text{add}$ (typically around 2000 determinants in these calculations). Single determinant projected energy estimates are accumulated for this equilibrated semi-stochastic walker dynamics until small random errors about a mean value are obtained. Each point took approximately 48 hours on a single 20 core compute node, representing calculations far from the limit of FCIQMC applicability given the effective parallelism of the method\cite{doi:10.1080/00268976.2013.877165,doi:10.1063/5.0005754}.

\begin{figure}[t]
\centering
\includegraphics[width=0.5\textwidth]{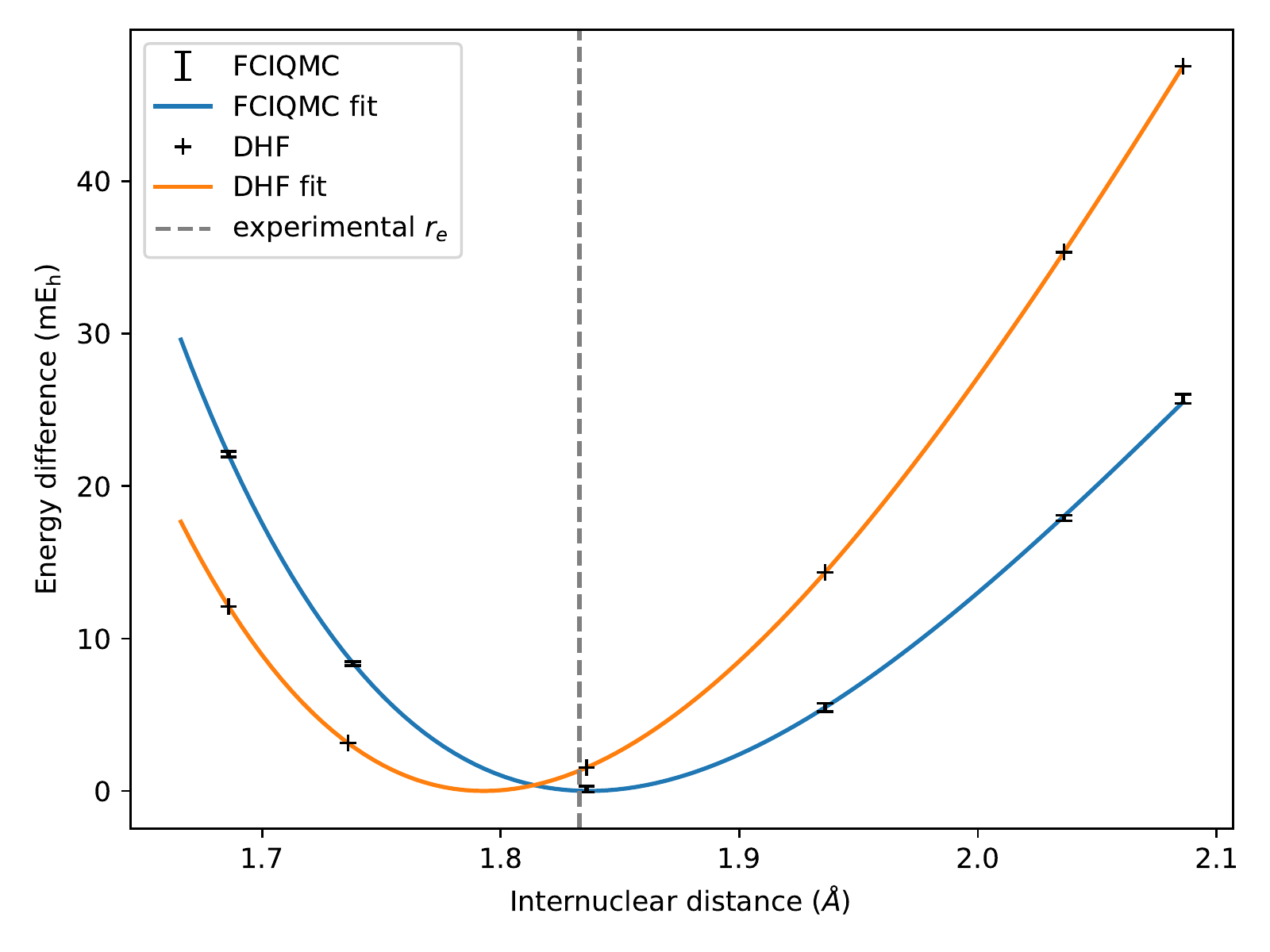}
	\caption{Dirac Hartree--Fock and 4c-FCIQMC energies in a KRDHF basis of 28 electrons and 122 Kramers paired spinors for SnO, shifted such that their equilibrium is at zero energy. Random errors are shown on the plot. Also shown is a fit to a simple Morse potential for each set of points, and the experimental\cite{doi:10.1063/1.1674155, rh16} value for $r_e$.
	}
\label{fig:SnO_morse_fit}
\end{figure}

The results of the 4c-FCIQMC study are shown in Fig.~\ref{fig:SnO_morse_fit}, where DHF and four-component FCIQMC results are fit to a simple Morse potential. The DHF potential gives a significantly shorter bond length of 179.3~pm compared to FCIQMC at 183.7~pm, and a substantially stiffer bond, with harmonic constant of $964 \pm 2$~cm$^{-1}$, compared to the FCIQMC value of $852\pm 30$~cm$^{-1}$, where the error represents the uncertainty in the fit. The FCIQMC values agree well with experiment\cite{doi:10.1063/1.1674155}, where it is found that $r_e=183.3$~pm, and $\omega_e=815$~cm$^{-1}$. Furthermore, agreement is also good with a previous CCSD(T) study which used a Douglas--Kroll--Hess Hamiltonian \cite{doi:10.1063/1.1690757} which was sufficient to obtain spectroscopic constants of $r_e=183.6$~pm and $\omega_e=812.3$~cm$^{-1}$.

Remaining discrepancies in the $\omega_e$ value are likely due to an insufficient active space, and the stochastic errors resulting in a wider range of potential fits to the Morse form. An important development to mitigate these effects will be to adapt the four-component FCIQMC algorithm for use with improved trial wave functions for the projected energy estimator, which can dramatically reduce random errors and has yet to be implemented within the four-component framework \cite{doi:10.1063/1.4920975}. Orbital optimization -- routine now in non-relativistic FCIQMC calculations\cite{doi:10.1021/acs.jctc.5b00917, doi:10.1021/acs.jctc.5b01190} -- can also be combined with the relativistic framework to self-consistently improve the choice of correlated orbital space. Furthermore, the use of internally-contracted multireference methods can also include a perturbative coupling to high-energy excitations \cite{doi:10.1063/1.5140086,doi:10.1080/00268976.2020.1802072}.



\section{Conclusion}

In this work we demonstrate the viability and efficiency of an FCIQMC framework for correlated electron problems with heavy elements. The well-demonstrated strengths of FCIQMC in a non-relativistic context transfer well over to the four-component adaptation, where the exploitation of sparsity in the spinor-basis wave function can overcome many of the traditional difficulties of four-component approaches, such as the increase in Hilbert space dimension, and loss of symmetries. Correlated problems far beyond the deterministically tractable number of electrons and basis spinors can be met, without a significant increase in computational effort compared to their non-relativistic analogue.

Consideration is given to the inclusion of point-group and time-reversal symmetries, both in the FCIQMC dynamic, and the adaptation of the underlying single-particle basis. Initial application to Tin Oxide demonstrates an accurate and effective approach, with significant potential to scale up to larger systems. From this feasibility study, it is appears worthwhile to now generalise other adaptations and efficiency improvements of non-relativistic FCIQMC for use in the four-component domain, including computation of reduced density matrices\cite{doi:10.1063/1.4904313, doi:10.1063/1.4986963}, trial wavefunctions for improved stochastic errors\cite{doi:10.1063/1.4920975}, sampling of excited states for zero-field splittings\cite{doi:10.1063/1.4932595}, adaptive shift to improve initiator error\cite{doi:10.1063/1.5134006}, and use within multireference perturbation theories\cite{doi:10.1063/1.5140086,doi:10.1080/00268976.2020.1802072}.


\section*{Data availability}
The data that support the findings of this study are available from the corresponding author upon reasonable request.

\begin{acknowledgements} 
The authors would like to sincerely thank Toru Shiozaki and Stefan Knecht for discussions and valuable technical help in obtaining the relevant integrals. G.H.B. gratefully acknowledges support from the Royal Society via a University Research Fellowship, as well as funding from the European Union's Horizon 2020 research and innovation programme under grant agreement No. 759063. We are also grateful to the UK Materials and Molecular Modelling Hub for computational resources, which is partially funded by EPSRC (EP/P020194/1).
\end{acknowledgements}

%
\end{document}